\documentclass[conference]{IEEEtran}
\IEEEoverridecommandlockouts
\usepackage{cite}
\usepackage{amsmath,amssymb,amsfonts}
\usepackage{algorithmic}
\usepackage{graphicx}
\usepackage{textcomp}
\usepackage{xcolor}
\usepackage{multirow}
\def\BibTeX{{\rm B\kern-.05em{\sc i\kern-.025em b}\kern-.08em
    T\kern-.1667em\lower.7ex\hbox{E}\kern-.125emX}}
\begin{document}

\title{A Modulation Front-End for Music Audio Tagging\\

\thanks{Cyrus Vahidi is a researcher at the UKRI CDT in AI and Music, supported jointly by the UKRI [grant number EP/S022694/1] and Music Tribe.}
}

\author{\IEEEauthorblockN{Cyrus Vahidi, Charalampos Saitis, Gy\"orgy~Fazekas}
\IEEEauthorblockA{\textit{Centre for Digital Music} \\
\textit{Queen Mary University of London}\\
London, United Kingdom \\
\{c.vahidi, c.saitis, g.fazekas\}@qmul.ac.uk}
}

\maketitle

\begin{abstract}
Convolutional Neural Networks have been extensively explored in the task of automatic music tagging. The problem can be approached by using either engineered time-frequency features or raw audio as input. Modulation filter bank representations that have been actively researched as a basis for timbre perception have the potential to facilitate the extraction of perceptually salient features. We explore end-to-end learned front-ends for audio representation learning, ModNet and SincModNet, that incorporate a temporal modulation processing block. The structure is effectively analogous to a modulation filter bank, where the FIR filter center frequencies are learned in a data-driven manner. The expectation is that a perceptually motivated filter bank can provide a useful representation for identifying music features. Our experimental results provide a fully visualisable and interpretable front-end temporal modulation decomposition of raw audio. We evaluate the performance of our model against the state-of-the-art of music tagging on the MagnaTagATune dataset. We analyse the impact on performance for particular tags when time-frequency bands are subsampled by the modulation filters at a progressively reduced rate. We demonstrate that modulation filtering provides promising results for music tagging and feature representation, without using extensive musical domain knowledge in the design of this front-end.
\end{abstract}

% \begin{IEEEkeywords}
% audio representation learning, automatic music tagging, auditory
% \end{IEEEkeywords}

\section{Introduction}\label{intro}
Deep learning techniques have emerged as a dominant force in modelling music and audio for retrieval and synthesis purposes, with a departure from feature engineering \cite{humphrey2013feature}. Mel spectrograms remain a popular feature representation in music information retrieval (MIR) among several tasks \cite{choi2018comparison, purwins2019deep} due to the robustness and compactness of the representation that enables a reduced data size and computational overhead. In MIR, music is viewed as a hierarchical composition of features and concepts at multiple levels of abstraction \cite{humphrey2013feature}. The feature extraction approach, that is common in MIR, typically consists of a cascade of fixed operations consisting of an affine transform, pooling and nonlinearities. This is analogous to the topology of a Convolutional Neural Network (CNN), where the operations' parameters are learnable. 

Mel filter banks, however, use a fixed basis. End-to-end Deep Neural Networks (DNNs) provide capacity to learn feature extractors that are highly optimised for the task at hand. End-to-end learned filter banks can be designed via DNNs, through the use of inductive biases towards the type of signal being modelled, for example, a layer of 1-D convolution FIR filters. End-to-end filter bank learning has been an active field of research in automatic speech recognition (ASR) \cite{sainath2015learning, cakir2016filterbank, ravanelli2018speaker, zeghidour2018learning, zeghidour2021leaf} and speech separation \cite{pariente2020filterbank}. This has demonstrated that randomly initialized 1-D convolutional kernels can learn a filter bank representation that performs time-frequency (TF) analysis under various parameterisations. Auditory domain knowledge can give insight into how to design an end-to-end learned filter bank, such that they align with representations known to pertain to human auditory perception. For example, the human auditory system is known to process the temporal envelope of critical bands via a modulation filter bank \cite{sek2003testing}.

In music tagging, Dieleman et al. \cite{dieleman2014end} first demonstrated that end-to-end automatic music taggers learn a time-frequency decomposition at the initial layers. Closer to current state-of-the-art, Musicnn \cite{pons2017end} applies musical domain knowledge in the design of its 2-D convolution kernel shapes, i.e. the various spectral profiles and temporal dependencies that arise from timbre and musical rhythmic structure. SampleCNN departs from the frame-based approach that is common when processing raw audio and STFT features, through the use of very small convolutional filters applied directly to raw waveform \cite{lee2018samplecnn}. This assumes that fine-grained local features can be hierarchically combined into a rich, structured representation. More recently, the HarmonicCNN demonstrated the utility of exploiting harmonic relationships in the front-end of a CNN music tagger \cite{won2020data}. HarmonicCNN utilises the auditory and musical domain knowledge that stipulates harmonic structure as key to human perception of pitch and tonality, and is inherent to western music that is based on an equal-tempered scale. Of the aforementioned models, a simple CNN with mel input features was shown to perform best in auto-tagging on the MagnaTagATune dataset in \cite{won2020evaluation}, yet less robust to generalization on perturbed inputs than the HarmonicCNN model. None of these models consider perceptually motivated representations for timbre that arise in auditory experiments.

A range of TF decomposition \emph{front-ends} have been learned for raw waveform modelling. This has been achieved through 1-D convolution and features of well-known auditory and signal processing filters, including Gamma-tone filters \cite{sainath2015learning}, sinc band-pass filters (SincNet) \cite{ravanelli2018speaker} and Gabor filters \cite{zeghidour2021leaf}. In ASR, multiscale feature learning has been explored by using variable-length TF kernels, where the kernel size biases the network to focus on appropriate frequency ranges \cite{zhu2016learning}. A multiresolution DNN front-end for ASR has been implemented using a 1D temporal convolution TF decomposition, followed by an \emph{envelope extractor} in place of the max-pooling operation, enabling varying resolution subsampling of learned critical band outputs \cite{tuske2018acoustic}. Most recently, an audio feature extraction front-end, LEAF \cite{zeghidour2021leaf}, that adopts learnable gabor filters, low-pass pooling and compression, marginally outperformed time-domain filter banks, SincNet and mel features on several audio classification tasks. The tasks included speaker identification, audio event detection and musical instrument recognition, however, automatic music tagging was not explored. 

Few end-to-end music tagging models make extensive use of perceptual domain knowledge in architecture design. Simulation of cochlear and cortical processing consists of a two-stage process involving a constant-Q filter bank followed by spectro-temporal modulation filtering \cite{chi2005multiresolution}. In the domain of auditory cognition,  modulation representations are actively researched for effectiveness as a vehicle for sound source identification and timbre perception \cite{patil2012music, thoret2017perceptually, elhilali2019modulation}. Dau's auditory excitation model consists of a cochlea gamma-tone ERB-scale filter bank, followed by half-wave rectification, low-pass processing of critical band outputs, and a modulation filter bank that performs analysis across filter band outputs \cite{dau1997modeling}. The model assumes that the temporal envelope within each critical band is processed by a bank of modulation filters \cite{sek2003testing}.

In this work, we investigate end-to-end learned front-end audio representations that are inspired by models of auditory modulation processing. The main contribution is a convolutional temporal modulation extractor that acts as a time-averaging operation. This layer enables a multiresolution temporal processing of filter band outputs. We evaluate the architecture on the task of automatic music tagging on the MagnaTagATune dataset. This is inherently related to the task of timbre analysis \cite{pons2017timbre}, a musical feature that gives rise to the semantic percepts of instrumentation and genre. Through analysis and visualisation of the learned weights we show that such an architecture can provide useful signal analysis and learning capacity. We compare the use of different network parameters and types of filter used for temporal modulation processing, perform a tag-based analysis of the network and contrast its performance against the state-of-the-art in music tagging on the MagnaTagATune dataset.

%%%%%%%%%%%%%%%%%%%%%%%%%%% METHOD AND ARCHITECTURE
\begin{figure*}[!htb]
\centerline{\includegraphics[width=\textwidth]{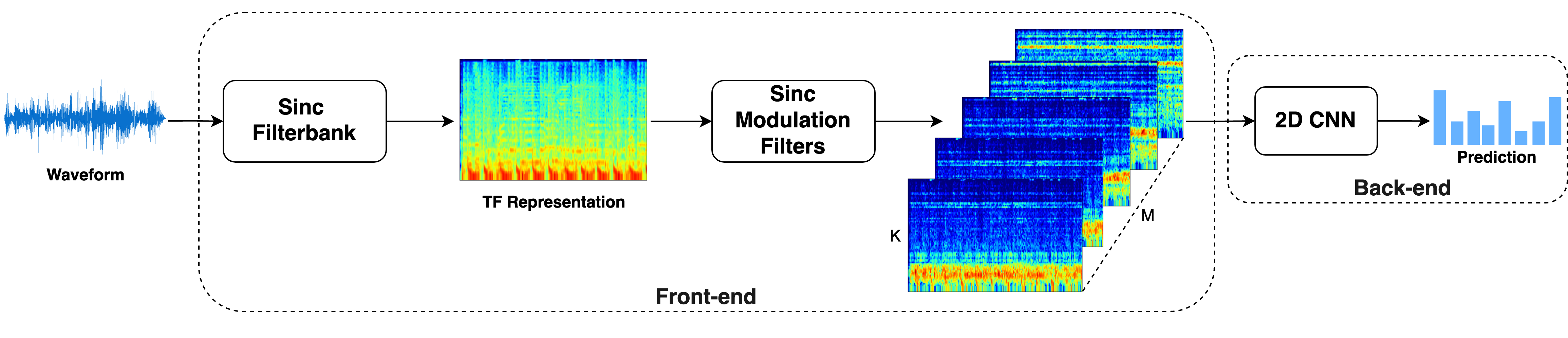}}
\caption{The proposed architecture. 1-D Modulation filters are applied across frequency bands of the time-frequency (TF) representation. A 3-D representation, where each channel represents a different modulation rate, is passed to a 2-D CNN classifier.}
\label{arch}
\end{figure*}

\section{System Overview}

\subsection{Modulation Front-end} \label{front-end}

\subsubsection{Learnable Time-Frequency Decomposition}
Commonly, the first layer of time-domain CNNs represents a set of finite impulse response (FIR) filters that perform a TF analysis. An inherent issue with learning filter banks through randomly initialised 1-D convolution is the non-optimality of the learned basis and the lack of constraints on the center frequencies during training. SincNet \cite{ravanelli2018speaker} facilitates learning filter banks by fixing the front-end TF analysis as a set of sinc band-pass filters, where the learnable parameters are the filters' cut-off frequencies. In place of 1-D convolutional kernels, we use a sinc band-pass time-frequency decomposition, with center frequencies initialized on the mel-scale. This choice is due to (a) constrained parameterization, (b) reduced number of parameters and (c) ease of interpretation of the learned set of filter parameters. In practice, we found the sinc filter bank to be more effective for both convergence stability and performance.

Equation \eqref{sinc} shows the convolution operation between the waveform ($x$) and sinc filter $g_k$, $k \in [1 ... K]$. Each filter's learnable parameters, $\theta_k$, consist of a low ($f_1$) and high ($f_2$) cut-off frequency. Equation \eqref{sinc2} is the formulation of each band-pass filter in the time-domain. This forms the basis for time-frequency decomposition in the front-end of the network. In our experiments, each kernel has a length of 256 samples (16 ms) with a convolution stride of 10 samples (0.625ms) at sampling rates of 16000 Hz. This can be followed by the ReLU or squared-modulus nonlinearity for half-wave rectification.

\begin{equation}\label{sinc}
    y_{k}[n] = x[n] * g_k[n, \theta_k]
\end{equation}

\begin{equation}\label{sinc2}
\begin{aligned}
    g_{k}[n, f_{1,k}, f_{2, k}] = 2 f_{2, k} sinc(2 \pi f_{2, k} n) - \\ 2 f_{1, k} sinc(2 \pi f_{1, k} n)
\end{aligned}
\end{equation}

% \subsubsection{low-pass filtering}
% TBD

\subsubsection{Temporal Modulation Processing}
In ASR, it is common to use a time-averaging layer, such as maximum pooling \cite{ravanelli2018speaker} or low-pass pooling \cite{zeghidour2018learning, zeghidour2021leaf}, across the filter bands of a filter bank output. In the layer that follows the TF analysis, we define a TF band time-averaging operation via a learnable set of FIR filters that are shared across incoming frequency channels. This formulation can enable learning of envelopes across the temporal dimension of the TF representation at varying resolutions. This can be viewed as a modulation filter bank if appropriate band-pass filter shapes can be learned. We use up to 20 of these envelope extractors.

We explore two types of envelope extraction: (a) fully learnable 1-D convolution FIR filters (ModNet), and (b) a bank of sinc band-pass filters (SincModNet). Both are applied to the time-frequency representation, with the filters shared across incoming frequency bands. In the case of learnable FIR filters, we initialise each FIR filter in the set as a Hamming window in one of 5 different positions across the filter, as in \cite{tuske2018acoustic}. For SincModNet, the band-pass filters are initialized on a linear scale. The convolution is strided by 160 samples (10ms) on the input that was downsampled to 1600 Hz by the TF analysis filter bank. Equation \eqref{layer2} expresses the operation of this layer, where $s_{m, k}$ is the output from modulation filter $h_m$ being applied across filter band $k$. This yields a 3-D representation from the $M$ modulation filters that applied across time and shared in frequency for each of the $K$ band-pass outputs.

\begin{equation}\label{layer2}
    s_{m, k}[n'] = y_{k}[n'] * h_{m}[n', \theta_m]
\end{equation}

An issue with the unconstrained optimisation of FIR filter weights is that they must learn both frequency bands to filter and a magnitude scaling factor, as was recognised in \cite{zeghidour2021leaf}. This was problematic during training, since some filters were able to learn meaningful filter shapes but displayed very low amplitude scaling. To effectively solve this issue, we utilised instance normalization \cite{ulyanov2016instance} that was shown to be effective in ASR \cite{zeghidour2018learning}. This applies mean-variance normalization per envelope channel and per audio example. This was found to be preferable to batch normalization for convergence and performance. For ModNet, this enabled meaningful filter shapes to be learned with more useful magnitude scaling factors. An alternative solution is L2 weight normalization, as was used in \cite{zeghidour2021leaf}.

\begin{table}[tbp]
    \centering
    \caption{Modulation Front-end Layers}
    \begin{tabular}{c|c}
        \hline  \hline
         Sinc convolution &  1-80-256-10 \\ \hline
         nonlinearity & $|.|^2$ or ReLU \\ \hline
         learnable FIR or sinc convolution & 80-20-128-160 \\ \hline
         nonlinearity & $|.|^2$ or ReLU \\ \hline
         normalization & instance or weight normalization\\ \hline
    \end{tabular}
    \label{tab:filter bank_layers}
\end{table}

\subsection{Back-end}
Deep networks for audio representation can be viewed from the perspective of a front-end feature extractor and a back-end classifier. The 3-D front-end representation is subsequently analysed by a 2-D CNN that is kept constant. Each of the modulation filter outputs is treated as an input channel into the 2-D CNN. The back-end is a ResNet architecture similar to that of HarmonicCNN \cite{won2020data}, consisting of seven 2D convolutional layers. Each layer consists of 256 channels, batch normalization \cite{ioffe2015batch} and ReLU nonlinearity. The final layer is a densely connected layer, with the sigmoid activation function for performing the task of multi-label music tagging.

\subsection{Implementation Details}
All models were trained for 200 epochs. The best model is was selected according to an early stopping regularizer with a patience of 15 epochs. A learning rate of $10^{-3}$ was used with a batch size of 4. Adam optimisation \cite{kingma2014adam} was used with a learning rate scheduler that halves the learning rate after observing no decrease in the validation loss within 5 epochs.  The audio sampling rate was fixed to 16 kHz and the audio input size 5 to seconds. All models were trained on RTX 2080 GPUs and took approximately 1 day to train. The code for all experiments is made available on Github\footnote{https://github.com/rastegah/modnet.git}.

\section{Datasets and Tasks}
We focus on automatic music tagging - the task of multi-label classification - where the presence of multiple semantic tags are predicted for a music clip. We experimented with a subset of the widely used MagnaTagATune (MTAT) dataset \cite{law2009evaluation} of around 26k audio clips. Each audio clip is around 30 seconds long. Only the top 50 tags are used, resulting in around around 21k audio samples for our experiments. We follow a split of 16 folds that has been used in many previous studies \cite{won2020evaluation}, where the first 12 folds are used for training, the 13th for validation and the remainder for testing. Many music tags that are based on mood, instrumentation and genre are highly related to the timbral characteristics of the audio.

\section{Experimental Results}
In this section, we present our experimental results for ModNet and SincModNet. Section \ref{param-study} compares various parameter configurations of the model. Section \ref{tag-based} compares performances across tags and interprets the effects of varying the modulation filter bank strides on tag-based performances. Sections \ref{learned-filters} and \ref{viz-exp} look more closely at the learned filters and their visualisation in the cases of both ModNet and SincModNet. Finally, in Section \ref{sota-compare}, we compare the performance of our model against state-of-the-art music taggers on the MTAT dataset.

\begin{figure*}[tb]
\centerline{\includegraphics[width=\textwidth]{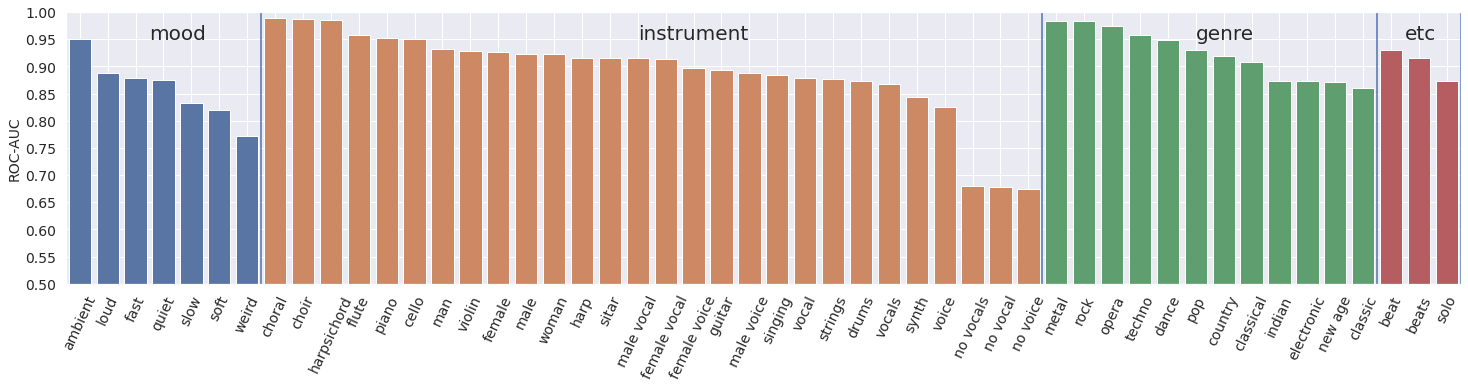}}
\caption{Tagwise performance grouped by category for the best performing SincModNet model using 20 modulation filters of length 128. ROC-AUC is reported per tag.}
\label{tagwise_perf}
\end{figure*}

\subsection{Parameter Study} \label{param-study}
We compare our temporal modulation analysis operation against a maximum pooling baseline to assess whether this signal processing step and representation is valuable for the CNN classifier. Table \ref{tab:mp} compares the performance of the two models. We report results for a SincModNet model using a small number of FIR envelopes, i.e. 5, in order to demonstrate the utility of this time-averaging operation. For a small number of modulation filters, the operation is more effective. Between 0 and 800 Hz, two of the five modulation frequencies are learned below 20 Hz, another between 30-40 Hz, and the remaining two at around 175 and 325 Hz.
\begin{table}[htbp]
    \centering
    \caption{Max Pooling vs learnable depthwise sinc band-pass filters}
    \begin{tabular}{c c c c c c}
        \hline
        Kernel Size & Stride & Model & ROC-AUC & PR-AUC\\ \hline
        \multirow{2}{*}{128} & \multirow{2}{*}{128} & Max Pooling & 0.8599 & 0.3484  \\ 
        & & 5 Mod Filters & \textbf{0.8790} & \textbf{0.3717} \\ \hline
    \end{tabular}
    \\
    \label{tab:mp}
\end{table}

\begin{figure}[bp]
\centerline{\includegraphics[width=\linewidth]{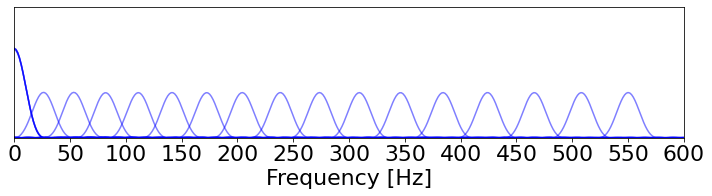}}
    \caption{Frequency response of the 20 learned modulation filters for a SincModNet model.}
\label{sinc-mod}
\end{figure}

Table \ref{tab:sincbpf} shows results for varying numbers of SincModNet modulation filters. We observe significant improvements across runs when 20 channels are used, indicating that the network prefers more granular subsampling rates of frequency bands. It should be noted that there is a slight drop in performance when moving from 5 to 15 filters. Upon inspection of the frequency response of the learned filters for the reported models, we observe that in the case of 15 filters, none of the center frequencies are below 20 Hz, while the 5 filter case does result in filters below 20 Hz. Rectifying this issue would require further stabilisation of training. The slower, sub audio-rate energy fluctuations are of particular interest in music processing e.g. tremolo and slower rhythmic structures that arise from articulation and musical structure \cite{FRAISSE1982149}. This also suggests the importance of initialization in the distribution of the resulting modulation filter frequencies. The frequency response of the 20 learned modulation filters are shown in Fig. \ref{sinc-mod}, showing that the resulting center frequencies span the important modulation range.

\begin{table}[tbp]
    \centering
    \caption{SincModNet Accuracy Comparison of number of modulation filters}
    \begin{tabular}{c | c c c}
        \hline
        Kernel Size & \# Modulation Filters & ROC-AUC & PR-AUC\\ \hline
        \multirow{4}{*}{128} & 5 & 0.8790 & 0.3717\\ \cline{2-4}
        & 10 & 0.8779 & 0.3684\\ \cline{2-4}
        & 15 & 0.8696 & 0.3551 \\ \cline{2-4}
        & 20 & \textbf{0.8939} & \textbf{0.4039} \\ \cline{2-4} \hline
    \end{tabular}
    \\
    \label{tab:sincbpf}
\end{table}

In Table \ref{tab:nonlinearity}, we compare the performance of different combinations of nonlinearities used in the front-end. We utilise a combination of ReLU and squared-modulus nonlinearities. The use of the absolute-squared nonlinearity brings the front-end representation closer to a power spectrum computed for various modulation rates. In practice, we find that a combination of ReLU and squared-modulus provides superior performance for this architecture. This aligns with the computation of a temporal modulation spectrum, where half-wave rectification can be used for filter-band envelope smoothing and the representation is transformed into power or energy magnitudes after modulation filtering \cite{elhilali2019modulation}.

\begin{table}[bp]
    \centering
    \caption{Performance comparison nonlinearity combinations after the TF analysis ($r1$) and modulation filtering ($r2$) layers.}
    \begin{tabular}{c c c}
        \hline
        $r1$ & $r2$ & ROC-AUC\\ \hline
        ReLU & $|.|^2$  & \textbf{0.8911} \\ \hline
        ReLU & ReLU & 0.8730\\ \hline
        -   & ReLU & 0.8 \\ \hline
        % -   & $|.|^2$ & - \\ \hline
    \end{tabular}
    \label{tab:nonlinearity}
\end{table}

\subsection{Tag-based performance} \label{tag-based}
Fig. \ref{tagwise_perf} displays the tag-wise ROC-AUC scores, grouped per tag category, in order to gain insight into whether there are significant differences between tags and tag categories. On comparison with the tag-based performance plots for HarmonicCNN and SampleCNN, we find that the performance trend among classes was clearly similar. This suggests that tag-based performance for this architecture is related to the notion of \emph{tagability} and class imbalances within music tagging data \cite{choi2018effects}. Since temporal modulation is known to characterise timbre features, varying the modulation filter parameters can give insight into how the filter bank affects performance on tags that are strongly related to timbre. Hence, to assess the impact of the modulation filters, we investigate tag-based performance under modulation rates at various levels of temporal coarseness by varying the stride of the modulation filters. 

\begin{figure}[tbp]
\centerline{\includegraphics[width=\linewidth]{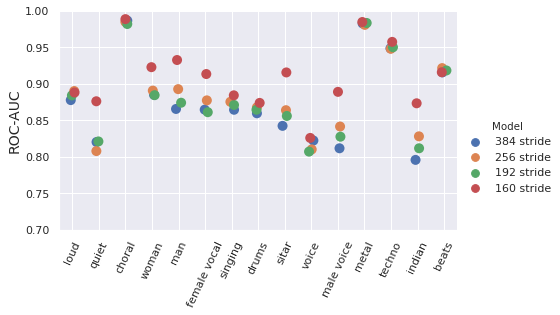}}
\caption{Comparison of the performance on selected tags for different convolutional stride (in samples) of the modulation filtering layer.}
\label{strides}
\end{figure}

Fig. \ref{strides} compares the effects on performance from varying the stride of the modulation filter bank. This intends to generate insight into the time-scales that modulations operate on for particular tags. The tag examples that are shown are those that were either mostly unaffected by reducing the modulation sampling rate, and those that showed a significant impact from increased stride. First, we draw attention to the contrasting impact of stride on the tags: \emph{loud} and \emph{quiet}. Analysing modulations at a finer resolution appears to be of significance in the case of \emph{quiet} examples. This alludes to a theory that transients are coded over coarser time-scales. In \cite{marinelli2020musical}, analysis of the salience of temporal modulations for dynamics classification supports the notion that temporal modulations for certain \emph{loud} sounds are coded over coarser scales. This may also explain why tags such as \emph{rock}, \emph{metal} and \emph{drums} are largely unaffected by increasing stride, since they are more characterised by the presence of loud transients at regular rhythmic intervals. From an auditory perspective, a sense of regular pulse is accounted for at rates of modulation below 8 Hz \cite{FRAISSE1982149}. We observe this in the visualisation, where for low temporal modulation rates, transients in the lower end of the spectrum show high energy. However, quiet sounds generally have a sparser presence in the spectrum and finer-scale loudness modulations. Further disparities can be observed with tags such as \emph{singing}, that are unaffected by increasing strides, and tags that account for the presence of vocals such as \emph{male voice}. Instances of the tag \emph{choral} tend to contain unison and longer sustained vocals. In contrast, \emph{male voice} contains more fine temporal structure and dynamics variation. Finally, instances of \emph{sitar} and \emph{indian} contain more irregular rhythmic content, sparser events and micro-structure; these tags are clearly related and are both affected by the coarser analysis of modulation content.

\begin{figure}[tbp!]
\centerline{\includegraphics[width=\linewidth]{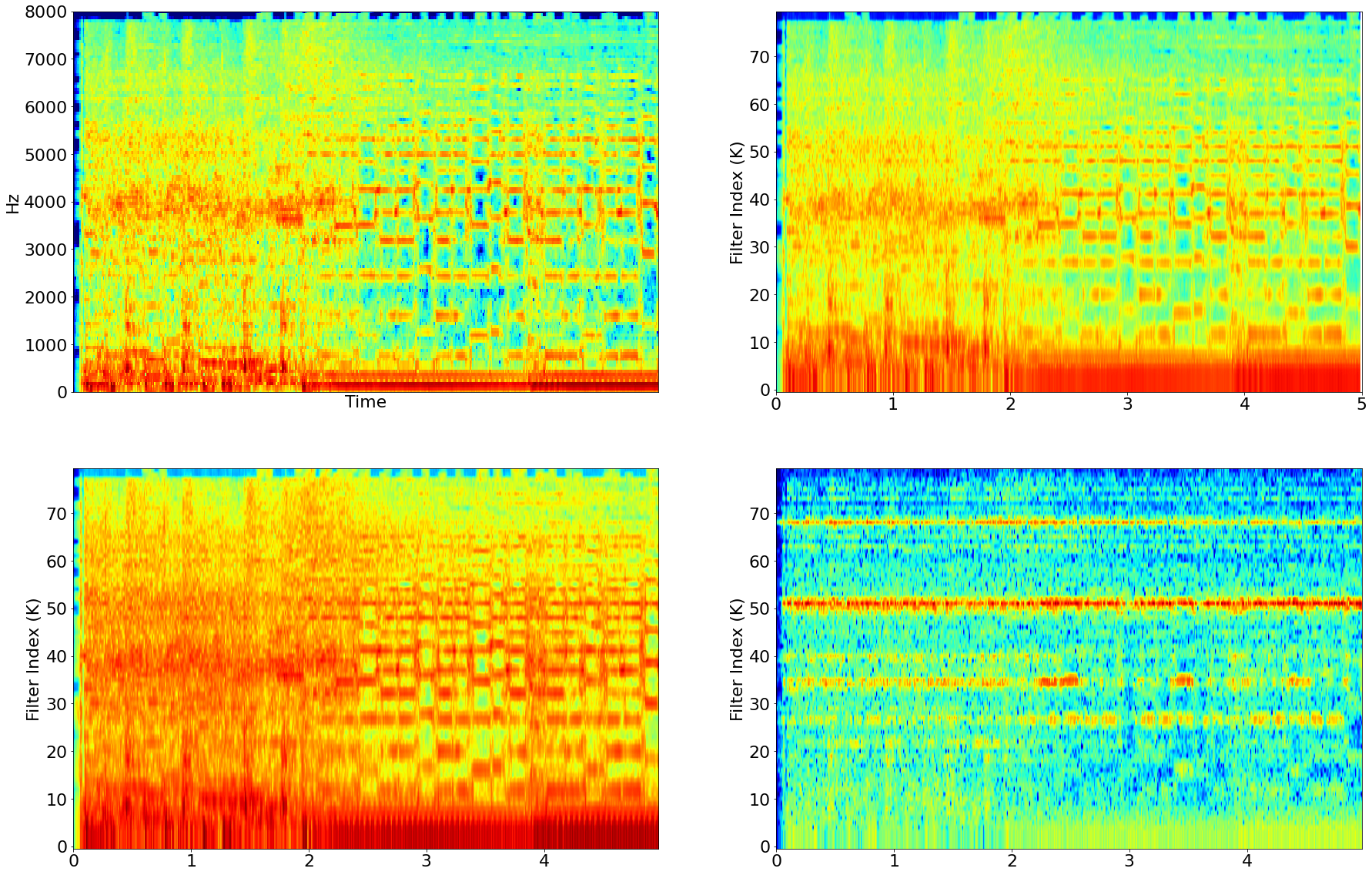}}
\caption{Magnitude (dB) spectrograms for (top-left) groundtruth (mel) (top-right) filter bank spectrogram (bottom-left) modulation spectrogram for the 1st modulation filter (bottom-right) modulation spectrogram for modulation filter 19. Each are from an example predominantly labelled \emph{rock}.}
\label{filter-audio}
\end{figure}
% // TODO: confusion matrix interpretation, strides tagwise perf comparison (time-scales)

\begin{figure}[tbp!]
\centerline{\includegraphics[width=\linewidth]{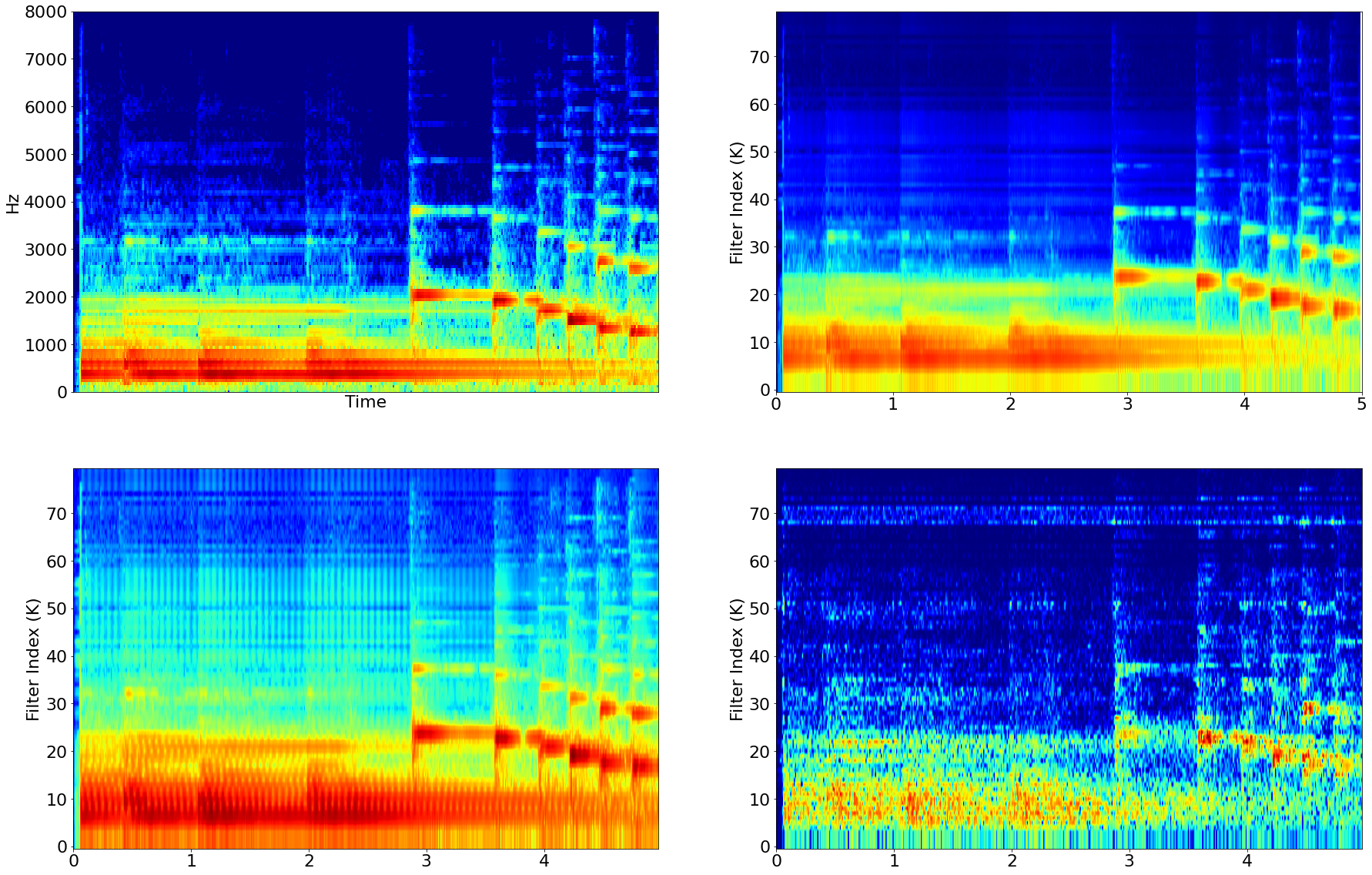}}
\caption{Magnitude (dB) spectrograms for (top-left) groundtruth (mel) (top-right) filter bank spectrogram (bottom-left) modulation spectrogram for the 1st modulation filter (bottom-right) modulation spectrogram for modulation filter 19. Each are from an example predominantly labelled \emph{quiet}.}
\label{filter-audio-2}
\end{figure}

\subsection{Learnt Modulation Filters} \label{learned-filters}
\begin{figure}[tbp]
\centerline{\includegraphics[width=\linewidth]{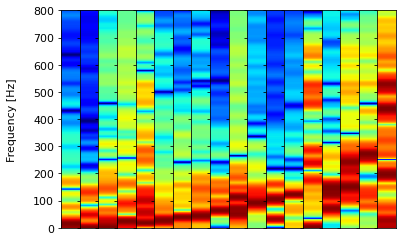}}
\caption{Learnt magnitude spectra of the 20 envelope extractors of a ModNet architecture (1-D convolution FIR filters of length 80), presented in a manually selected order. Although diverse filter shapes can be learnt e.g. low-pass and band-pass.}
\label{env-spec}
\end{figure}

Formulating the time-averaging layer as FIR filters makes interpretation possible. Fig. \ref{env-spec} shows the magnitude spectra for each of 20 filters learned with a ModNet architecture, where the filters are derived from randomly initialized 1-D convolution kernels. The layer has learnt various filter shapes that sample the TF bands at varying rates, predominantly within the range of 0 - 200 Hz. The filters tend to have close to low-pass or band-pass characteristics when the weights are learnt in a purely data-driven way, as was previously demonstrated with a similar architecture for ASR \cite{tuske2018acoustic}. Exemplary of the learnt low-pass filters, often the resulting impulse responses were clearly triangular shaped, but this result was subject to variation across training runs. These results incentivized the use of a constrained set of sinc band-pass modulation filters in SincModNet.

\begin{figure*}[t]
\centerline{\includegraphics[width=\textwidth]{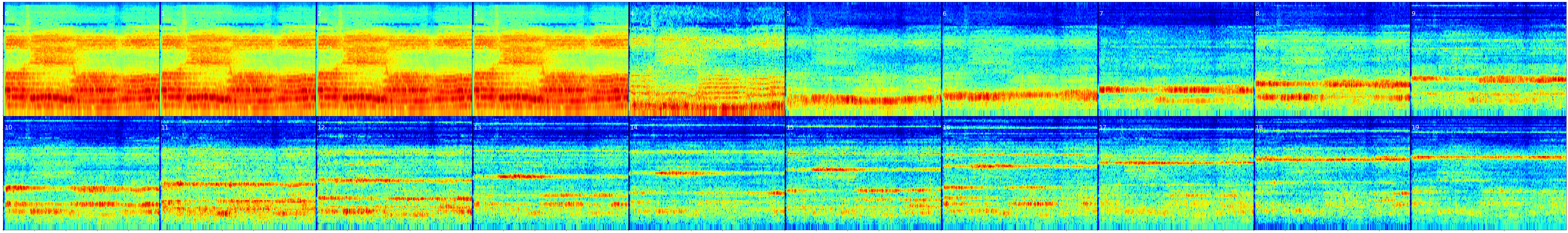}}
\caption{Visualisation matrix of the representation when modulation filters are applied to an instance of the \emph{choral} tag. Plotted in ascending order of modulation frequency from left to right. This gives an indication that modulation rates appear in differing frequency bands for the various tags.}
\label{viz_matrix}
\end{figure*}

\subsection{Visualisation of Learnt Representation} \label{viz-exp}
Inspection of the learnt representation for various tags can enable the identification of the types of amplitude modulations that the network prefers to prioritise, and in which frequency bands such temporal patterns lie. We can observe that modulations respond at different time-scales, in varying frequency bands between tags. Fig. \ref{filter-audio} and \ref{filter-audio-2} show localization of modulation content across frequency bands for the lowest and highest modulation rates. Fig. \ref{viz_matrix} illustrates the representation that results from applying the stack of modulation filters to the time-frequency input. Higher modulation rates generally show a sparser presence in the representation, since these occur over finer time scales. In the lowest, rhythmically perceived modulation ranges, i.e. below 8 Hz, percepts of meter and tremolo amplitude modulation can be detected.

\subsection{Comparison with state-of-the-art} \label{sota-compare}
\begin{table}[bp]
    \centering
    \caption{Comparison with SOTA}
    \begin{tabular}{c c c c}
        \hline
        & \multicolumn{2}{c}{Music Tagging (MTAT)} \\
        Model & ROC-AUC & PR-AUC\\ \hline
        Dieleman et al. \cite{dieleman2014end} & 84.87 & -  \\ \hline
        Musicnn \cite{pons2017end} & 0.9106 & 0.4493 \\ \hline
        SampleCNN \cite{lee2018samplecnn} & 0.9058 & 0.4422  \\ \hline
        HarmonicCNN \cite{won2019automatic} & \textbf{0.9127} & \textbf{0.4611} \\ \hline
        \textbf{ModNet} & \emph{0.8874} & \emph{0.3979}  \\ \hline
        \textbf{SincModNet} & \emph{0.8939} & \emph{0.4039}\\ 
    \end{tabular}
    \\
    \label{tab:sota}
\end{table}
We evaluate against the state-of-the-art models in music tagging on MTAT. Table \ref{tab:sota} shows area under the ROC curve (ROC-AUC) and the precision-recall curve (PR-AUC) as metrics for the predictive performance on the binary decision problem. These metrics are computed across individual tags and averaged. Our reported models are \emph{ModNet} (sinc band-pass filter bank and learnable FIR filter temporal envelope extraction) and \emph{SincModNet} (sinc band-pass filter bank and sinc band-pass temporal envelope extraction). Both use 80 TF filters of stride 0.625ms and 20 modulation filters of stride 10ms. ReLU ($r1$) and abs-squared nonlinearities ($r2$) are used for each layer respectively. We discussed the compared models in Section \ref{intro}. The results reported for the models are from \cite{won2020evaluation}, where further details of their implementation can be found. HarmonicCNN results are reported for the model that operates on raw waveform, introduced in \cite{won2019automatic}.

Our model performs marginally worse on ROC-AUC than HarmonicCNN and Musiccnn, although Musicnn operates on mel-spectrogram inputs. Compared to these models, our model does not employ musical domain knowledge, i.e. the use of harmonic filter stacks and musically motivated filter shapes. For this reason, the results are aligned with our expectations. Furthermore, our representation is open to further interpretation and engineering effort that could account better for the modulation content of the class of signals being modelled. Future work could adopt an approach that uses a fusion of musical domain knowledge, e.g harmonic filters combined with modulation processing. This would allow us to better discern the benefit of modulation processing in the front-end. In our model, aspects of music that are relevant to timbre are more emphasised, but these may be insufficient to boost the performance across all tag categories.

\section{Conclusions and future work}
In this work, we introduced a time-frequency time-averaging operation for end-to-end audio representation learning. We interpret this layer as a modulation filter bank that captures temporal amplitude modulations at various rates, through the use of learnable FIR filters. This is a first step in music tagging to explore the use of modulation representations in a fully end-to-end context. We visualised the representation, which shows the varying presence of the modulation rates across tags. Analysis of changing the stride of the modulation filters showed a varying impact on performance, suggesting that the time-scale of modulations and subsampling rate is of relevance for identifying particular tags, while coarse analysis of modulation content is permissible for several tags. 

The modulation filter center frequencies are learned in a data-driven manner. However, the current models do not directly replicate the time-frequency resolution of psychoacoustic models, i.e. bandwidth and modulation frequency settings. Additional signal processing steps can be introduced, such as a channel-wise low-pass smoothing step that mimics cortical processing. Domain knowledge of amplitude modulation rates for the class of signals being modelled could be incorporated to further tune the settings of the learnable modulation filters. Further architectural design could incorporate analysis of energy fluctuations at multiple time-scales.

Learnable modulation filters applied directly to STFT inputs are yet to be explored. Future work will evaluate the architecture on additional music tagging datasets and test its capability of generalization to perturbed inputs. This model could be applied to other tasks such as acoustic scene analysis and instrument identification tasks. It is of interest to explore the use of such front-ends in a similarity metric for downstream generative tasks such as audio denoising. Overall, the use of modulation processing in a feature extraction front-end can provide competitive performance on tagging of the MTAT dataset, without using extensive musical domain knowledge in the architecture's design.

\section*{Acknowledgment}
Cyrus Vahidi is a researcher at the UKRI CDT in AI and Music, supported jointly by the UKRI [grant number EP/S022694/1] and Music Tribe Brands UK Ltd.

\bibliography{refs.bib}

\begin{thebibliography}{10}
\providecommand{\url}[1]{#1}
\csname url@samestyle\endcsname
\providecommand{\newblock}{\relax}
\providecommand{\bibinfo}[2]{#2}
\providecommand{\BIBentrySTDinterwordspacing}{\spaceskip=0pt\relax}
\providecommand{\BIBentryALTinterwordstretchfactor}{4}
\providecommand{\BIBentryALTinterwordspacing}{\spaceskip=\fontdimen2\font plus
\BIBentryALTinterwordstretchfactor\fontdimen3\font minus
  \fontdimen4\font\relax}
\providecommand{\BIBforeignlanguage}[2]{{%
\expandafter\ifx\csname l@#1\endcsname\relax
\typeout{** WARNING: IEEEtran.bst: No hyphenation pattern has been}%
\typeout{** loaded for the language `#1'. Using the pattern for}%
\typeout{** the default language instead.}%
\else
\language=\csname l@#1\endcsname
\fi
#2}}
\providecommand{\BIBdecl}{\relax}
\BIBdecl

\bibitem{humphrey2013feature}
E.~J. Humphrey, J.~P. Bello, and Y.~LeCun, ``Feature learning and deep
  architectures: New directions for music informatics,'' \emph{Journal of
  Intelligent Information Systems}, vol.~41, no.~3, pp. 461--481, 2013.

\bibitem{choi2018comparison}
K.~Choi, G.~Fazekas, M.~Sandler, and K.~Cho, ``A comparison of audio signal
  preprocessing methods for deep neural networks on music tagging,'' in
  \emph{2018 26th European Signal Processing Conference (EUSIPCO)}.\hskip 1em
  plus 0.5em minus 0.4em\relax IEEE, 2018, pp. 1870--1874.

\bibitem{purwins2019deep}
H.~Purwins, B.~Li, T.~Virtanen, J.~Schl{\"u}ter, S.-Y. Chang, and T.~Sainath,
  ``Deep learning for audio signal processing,'' \emph{IEEE Journal of Selected
  Topics in Signal Processing}, vol.~13, no.~2, pp. 206--219, 2019.

\bibitem{sainath2015learning}
T.~N. Sainath, R.~J. Weiss, A.~Senior, K.~W. Wilson, and O.~Vinyals, ``Learning
  the speech front-end with raw waveform {CLDNNs},'' in \emph{Sixteenth Annual
  Conference of the International Speech Communication Association}, 2015.

\bibitem{cakir2016filterbank}
E.~Cakir, E.~C. Ozan, and T.~Virtanen, ``Filterbank learning for deep neural
  network based polyphonic sound event detection,'' in \emph{2016 International
  Joint Conference on Neural Networks (IJCNN)}.\hskip 1em plus 0.5em minus
  0.4em\relax IEEE, 2016, pp. 3399--3406.

\bibitem{ravanelli2018speaker}
M.~Ravanelli and Y.~Bengio, ``Speaker recognition from raw waveform with
  {SincNet},'' in \emph{2018 IEEE Spoken Language Technology Workshop
  (SLT)}.\hskip 1em plus 0.5em minus 0.4em\relax IEEE, 2018, pp. 1021--1028.

\bibitem{zeghidour2018learning}
N.~Zeghidour, N.~Usunier, I.~Kokkinos, T.~Schaiz, G.~Synnaeve, and E.~Dupoux,
  ``Learning filterbanks from raw speech for phone recognition,'' in \emph{2018
  IEEE international conference on acoustics, speech and signal Processing
  (ICASSP)}.\hskip 1em plus 0.5em minus 0.4em\relax IEEE, 2018, pp. 5509--5513.

\bibitem{zeghidour2021leaf}
N.~Zeghidour, O.~Teboul, F.~d.~C. Quitry, and M.~Tagliasacchi, ``{LEAF}: A
  learnable frontend for audio classification,'' \emph{arXiv preprint
  arXiv:2101.08596}, 2021.

\bibitem{pariente2020filterbank}
M.~Pariente, S.~Cornell, A.~Deleforge, and E.~Vincent, ``Filterbank design for
  end-to-end speech separation,'' in \emph{ICASSP 2020-2020 IEEE International
  Conference on Acoustics, Speech and Signal Processing (ICASSP)}.\hskip 1em
  plus 0.5em minus 0.4em\relax IEEE, 2020, pp. 6364--6368.

\bibitem{sek2003testing}
A.~Sek and B.~C. Moore, ``Testing the concept of a modulation filter bank: The
  audibility of component modulation and detection of phase change in
  three-component modulators,'' \emph{The Journal of the Acoustical Society of
  America}, vol. 113, no.~5, pp. 2801--2811, 2003.

\bibitem{dieleman2014end}
S.~Dieleman and B.~Schrauwen, ``End-to-end learning for music audio,'' in
  \emph{2014 IEEE International Conference on Acoustics, Speech and Signal
  Processing (ICASSP)}.\hskip 1em plus 0.5em minus 0.4em\relax IEEE, 2014, pp.
  6964--6968.

\bibitem{pons2017end}
J.~Pons, O.~Nieto, M.~Prockup, E.~Schmidt, A.~Ehmann, and X.~Serra,
  ``End-to-end learning for music audio tagging at scale,'' \emph{arXiv
  preprint arXiv:1711.02520}, 2017.

\bibitem{lee2018samplecnn}
J.~Lee, J.~Park, K.~L. Kim, and J.~Nam, ``Sample{CNN}: End-to-end deep
  convolutional neural networks using very small filters for music
  classification,'' \emph{Applied Sciences}, vol.~8, no.~1, p. 150, 2018.

\bibitem{won2020data}
M.~Won, S.~Chun, O.~Nieto, and X.~Serrc, ``Data-driven harmonic filters for
  audio representation learning,'' in \emph{ICASSP 2020-2020 IEEE International
  Conference on Acoustics, Speech and Signal Processing (ICASSP)}.\hskip 1em
  plus 0.5em minus 0.4em\relax IEEE, 2020, pp. 536--540.

\bibitem{won2020evaluation}
M.~Won, A.~Ferraro, D.~Bogdanov, and X.~Serra, ``Evaluation of {CNN}-based
  automatic music tagging models,'' \emph{arXiv preprint arXiv:2006.00751},
  2020.

\bibitem{zhu2016learning}
Z.~Zhu, J.~H. Engel, and A.~Hannun, ``Learning multiscale features directly
  from waveforms,'' \emph{arXiv preprint arXiv:1603.09509}, 2016.

\bibitem{tuske2018acoustic}
Z.~T{\"u}ske, R.~Schl{\"u}ter, and H.~Ney, ``Acoustic modeling of speech
  waveform based on multi-resolution, neural network signal processing,'' in
  \emph{2018 IEEE International Conference on Acoustics, Speech and Signal
  Processing (ICASSP)}.\hskip 1em plus 0.5em minus 0.4em\relax IEEE, 2018, pp.
  4859--4863.

\bibitem{chi2005multiresolution}
T.~Chi, P.~Ru, and S.~A. Shamma, ``Multiresolution spectrotemporal analysis of
  complex sounds,'' \emph{The Journal of the Acoustical Society of America},
  vol. 118, no.~2, pp. 887--906, 2005.

\bibitem{patil2012music}
K.~Patil, D.~Pressnitzer, S.~Shamma, and M.~Elhilali, ``Music in our ears: the
  biological bases of musical timbre perception,'' \emph{PLoS Comput Biol},
  vol.~8, no.~11, p. e1002759, 2012.

\bibitem{thoret2017perceptually}
E.~Thoret, P.~Depalle, and S.~McAdams, ``Perceptually salient regions of the
  modulation power spectrum for musical instrument identification,''
  \emph{Frontiers in psychology}, vol.~8, p. 587, 2017.

\bibitem{elhilali2019modulation}
M.~Elhilali, ``Modulation representations for speech and music,'' in
  \emph{Timbre: Acoustics, perception, and cognition}, K.~Siedenburg,
  C.~Saitis, S.~McAdams, A.~N. Popper, and R.~R. Fay, Eds.\hskip 1em plus 0.5em
  minus 0.4em\relax Springer, 2019, pp. 335--359.

\bibitem{dau1997modeling}
T.~Dau, B.~Kollmeier, and A.~Kohlrausch, ``Modeling auditory processing of
  amplitude modulation. i. detection and masking with narrow-band carriers,''
  \emph{The Journal of the Acoustical Society of America}, vol. 102, no.~5, pp.
  2892--2905, 1997.

\bibitem{pons2017timbre}
J.~Pons, O.~Slizovskaia, R.~Gong, E.~G{\'o}mez, and X.~Serra, ``Timbre analysis
  of music audio signals with convolutional neural networks,'' in \emph{2017
  25th European Signal Processing Conference (EUSIPCO)}.\hskip 1em plus 0.5em
  minus 0.4em\relax IEEE, 2017, pp. 2744--2748.

\bibitem{ulyanov2016instance}
D.~Ulyanov, A.~Vedaldi, and V.~Lempitsky, ``Instance normalization: The missing
  ingredient for fast stylization,'' \emph{arXiv preprint arXiv:1607.08022},
  2016.

\bibitem{ioffe2015batch}
S.~Ioffe and C.~Szegedy, ``Batch normalization: Accelerating deep network
  training by reducing internal covariate shift,'' \emph{arXiv preprint
  arXiv:1502.03167}, 2015.

\bibitem{kingma2014adam}
D.~P. Kingma and J.~Ba, ``Adam: A method for stochastic optimization,''
  \emph{arXiv preprint arXiv:1412.6980}, 2014.

\bibitem{law2009evaluation}
E.~Law, K.~West, M.~I. Mandel, M.~Bay, and J.~S. Downie, ``Evaluation of
  algorithms using games: The case of music tagging.'' in \emph{ISMIR}, 2009,
  pp. 387--392.

\bibitem{FRAISSE1982149}
P.~Fraisse, ``Rhythm and tempo,'' in \emph{Psychology of Music}, D.~Deutsch,
  Ed.\hskip 1em plus 0.5em minus 0.4em\relax San Diego: Academic Press, 1982,
  pp. 149--180.

\bibitem{choi2018effects}
K.~Choi, G.~Fazekas, K.~Cho, and M.~Sandler, ``The effects of noisy labels on
  deep convolutional neural networks for music tagging,'' \emph{IEEE
  Transactions on Emerging Topics in Computational Intelligence}, vol.~2,
  no.~2, pp. 139--149, 2018.

\bibitem{marinelli2020musical}
L.~Marinelli, A.~Lykartsis, S.~Weinzierl, and C.~Saitis, ``Musical dynamics
  classification with {CNN} and modulation spectra,'' 2020.

\bibitem{won2019automatic}
M.~Won, S.~Chun, O.~Nieto~Caballero, and X.~Serra, ``Automatic music tagging
  with harmonic {CNN},'' 2019.

\end{thebibliography}
\bibliographystyle{IEEEtran}

\end{document}